\documentclass[10pt]{iopart}
\usepackage{iopams}
\usepackage{graphicx}
\usepackage{citesort}
\usepackage{relsize}
\usepackage{scalerel}
\newcommand{\rmr}{\mathrm{r}}
\newcommand{\rmin}{\mathrm{in}}
\newcommand{\rmout}{\mathrm{out}}
\newcommand{\tix}{\tilde{\bi{x}}}
\newcommand{\tidx}{\rmd\tilde{\bi{x}}}
\newcommand{\tib}{\tilde{\bi{B}}}
\newcommand{\tik}{\tilde{\bi{K}}}
\begin{document}
\title[]{Analytic solutions to the Maxwell-London equations and levitation force for a superconducting sphere in a quadrupole field}
\author{J Hofer and M Aspelmeyer}
\address{Vienna Center for Quantum Science and Technology (VCQ), Faculty of Physics, University of Vienna, Boltzmanngasse 5, Vienna A-1090, Austria}
\ead{joachim.hofer@univie.ac.at}
\begin{abstract}
Recent proposals suggest using magnetically trapped superconducting spheres in the Meissner state to create low-loss mechanical oscillators with long coherence times. In these proposals the derivation of the force on the superconducting sphere and the coupling to the sphere typically relies on a vanishing penetration depth $\lambda$ as well as a specific symmetry (i.e. restricting the position of the sphere to one axis) or heuristic methods (e.g. assigning an equivalent point magnetic dipole moment to the sphere).
In this paper we analytically solve the Maxwell-London equations with appropriate boundary conditions for a superconducting sphere in a quadrupole field. The analytic solutions provide the full field distribution for arbitrary $\lambda$ and for an arbitrary sphere position as well as the distribution of shielding currents within the sphere. We furthermore calculate the force acting on the sphere and the maximum field over the volume of the sphere. We show that for a certain range of $\lambda$ the maximum field experienced by the superconducting sphere is actually lower than it is for a non-magnetic sphere. 
\end{abstract}
\vspace{2pc}

\submitto{\SUST}
\maketitle
\ioptwocol

\section{Introduction}
The last decade has seen significant progress in achieving quantum control over solid state mechanical devices. The main idea is to exploit the available toolbox of quantum optics in both the optical and microwave domain by coupling mechanical motion to optical cavities or superconducting circuits \cite{RevModPhys.86.1391}\cite{POOT2012273}. Recent examples include the generation of non-classical states of motion \cite{O'Connell2010}\cite{Wollman952}\cite{Hong203} and even of quantum entanglement between different micro-mechanical systems \cite{Riedinger2018}\cite{Ockeloen-Korppi2018}. Several proposals have suggested that quasi-static magnetic levitation of superconductors in the Meissner state allows to further increase both system size and coherence time in such experiments, thereby not only improving the system performance but also enabling access to a completely new parameter regime of macroscopic quantum physics \cite{PhysRevLett.109.147205}\cite{PhysRevLett.109.147206}\cite{2058-9565-3-2-025001}. The requirements on the magnetic traps are similar to those of atom traps\cite{RevModPhys.79.235}, as in both cases a minimum in the magnetic field norm is necessary for levitation. Several trap configurations, such as the Anti-Helmholtz setup suggested in \cite{PhysRevLett.109.147205}, produce a (local) quadrupole field, i.e. a magnetic field of the form $\case{1}{2}\,b_{\mathrm{z}}\,\{x,\,y,\,-2\,z\}$,
where $b_z$ denotes the magnetic gradient along the $z$-axis. Coupling to the motion of the sphere is facilitated by placing a pickup loop in the proximity of the sphere. The flux through the pickup loop then depends on the position of the sphere. A detailed knowledge of the magnetic field distribution for arbitrary sphere positions is essential for a good understanding of both the trap dynamics and the coupling strength. 
However, a full analysis of the magnetic field distribution of such a configuration has not been carried out yet. The proposals referenced above provide results only for vanishing penetration depth ($\lambda=0$) and rely either on symmetry features, where the sphere is restricted to the $z$-axis or use heuristic methods such as approximating the sphere as a point dipole. The aim of this paper is to avoid these restrictions and provide analytic expressions for the magnetic field for arbitrary $\lambda$ and arbitrary sphere positions.

In section 2 we derive the magnetic field and the supercurrent on the surface of the sphere for the special case $\lambda=0$ by solving the Maxwell equation with the appropriate boundary condition; in section 3 we generalize these results for arbitrary $\lambda$ by solving the Maxwell-London equations with the appropriate boundary condition. 

In section 4 we derive the force acting on the sphere and the maximum field over the volume of the sphere when it is located at the origin of the quadrupole field. The latter is important because if the maximum field seen by the sphere surpasses a critical field $B_{c1}$ (dependent on the material), the sphere will no longer be in the Meissner state. 

Section 5 provides a brief summary and a discussion of the results.

For mathematical simplicity we choose the coordinate system such that the superconducting sphere of radius $R$ is at the origin and the center of the quadrupole field is displaced relative to the origin by $-\bi{\rmd x}=-\{\rmd x,\,\rmd y,\,\rmd z\}$. We use $\bi{x}=\{x,\,y,\,z\}$ for the position vector in Cartesian coordinates. As is conventionally done, we refer to the magnetic flux density $\bi{B}$ as the magnetic field. The applied quadrupole field thus takes the form 
\begin{equation}
\label{eq:quad1}
\bi{B}_0 = \case{1}{2}\,b_{\mathrm{z}}\,\{x+\rmd x,\,y+\rmd y,\,-2\,(z+\rmd z)\}.
\end{equation}
We will also use spherical coordinates $(r,\,\theta,\,\phi)$ and the corresponding basis vectors $\bi{e}_\rmr,\,\bi{e}_\mathrm{\theta},\,\bi{e}_\mathrm{\phi}$. Spherical harmonics $Y_n^m$ are understood to have the normalization $Y_n^m=\sqrt{\case{2n+1}{4\pi}\,\case{(n-m)!}{(n+m)!}}\,P_n^m(\cos\theta)\,\exp(\rmi\phi)$, where $P_n^m$ stands for the associated Legendre polynomials\cite{jackson_classical_1999}. The vector potential and magnetic field inside the sphere are denoted by $\bi{A}_\rmin$ and $\bi{B}_\rmin$, respectively, while $\bi{B}_\rmout$ is used for the field outside the sphere. There is no current outside the sphere, so we can use a scalar potential $\Phi$ such that $\bi{B}_\rmout=\bi{B}_0-\nabla\Phi$. Since physical solutions for the induced field density $-\nabla\Phi$ vanish at infinity, it follows from $\Delta\Phi=0$ that
\begin{equation}
\label{eq:scalpot1}
\Phi=\sum_{n=0}^{\infty}r^{-(n+1)}\sum_{m=-n}^n a_{n,m}\,Y_n^m,
\end{equation}
where the coefficients $a_{n,m}$ will be determined below. Components of a vector will be denoted by a superscript rather than a subscript, e.g. $\bi{B}_\rmin^\rmr=\bi{B}_\rmin\cdot\bi{e}_\rmr$.
We will use the Coulomb gauge $\nabla\bi{A}=0$ for any vector potential $\bi{A}$ throughout this paper. 
\begin{figure}[ht!]
\centering
\includegraphics[width=\linewidth]{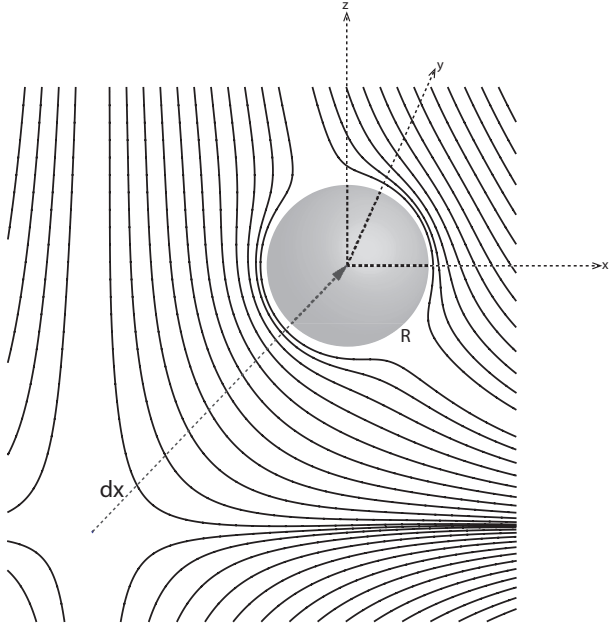}
\caption{Sketch of the geometry. The superconducting sphere, located at the origin of the coordinate system, is displaced from the center of the quadrupole field by $\bi{\rmd x}=\{\rmd x,\,\rmd y,\,\rmd z\}$. The field is depicted by its streamlines in the xz-plane.}
\label{fig1}
\end{figure}

\section{Magnetic field for $\lambda = 0$}
\begin{figure}[ht!]
\centering
\includegraphics[width=\linewidth]{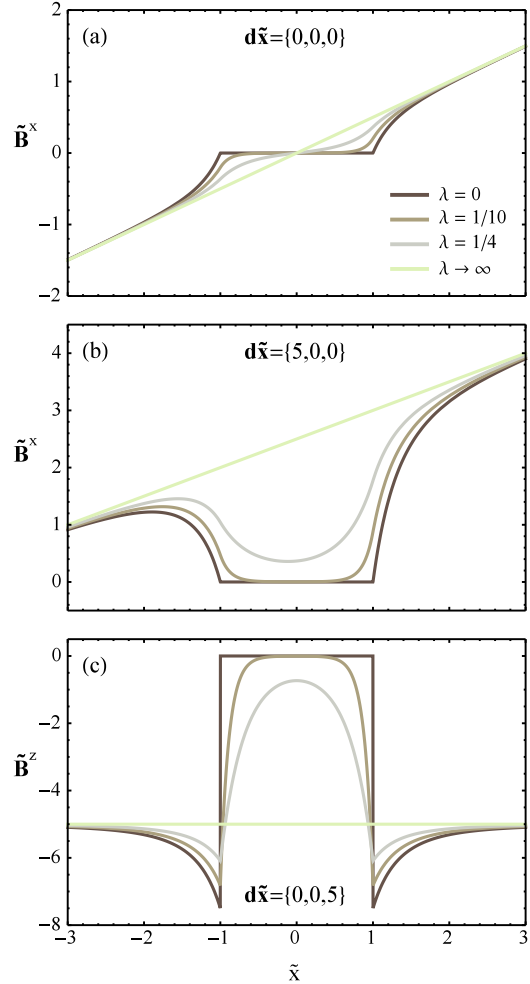}
\caption{Plot of normal and transverse magnetic field along the normalized x-axis for various penetration depths and sphere displacements. (a)Plot of normal magnetic field for $\tidx=\{0,\,0,\,0\}$. (b)Plot of normal magnetic field for $\tidx=\{5,\,0,\,0\}$. (c)Plot of transverse magnetic field for $\tidx=\{0,\,0,\,5\}$. The legend printed in panel (a) is valid for all panels.}
\label{fig2}
\end{figure}

For vanishing penetration depth there is no magnetic field inside the superconductor and the normal component of the magnetic field vanishes at the surface of the superconductor\cite{tinkham1996introduction}, which in our case corresponds to 
\begin{equation}
\label{eq:boundary1}
\bi{B}_\rmin=0,\,\,\bi{B}_\rmout^\rmr|_{_{r=R}}=0.
\end{equation}
It follows from \eref{eq:quad1} and \eref{eq:scalpot1} that the radial part of the applied field and the induced field are given by
\begin{eqnarray*}
\fl \bi{B}_0^\rmr = b_z \big(&-\sqrt{4\pi/3}\,\rmd z\,Y_1^0 + \sqrt{\pi/6}\,(\rmi\,\rmd y+\rmd x)\,Y_1^{-1} \\
&+ \sqrt{\pi/6}\,(\rmi\,\rmd y-\rmd x)\,Y_1^{1} -\sqrt{4\pi/5}\,r\,Y_2^0\big),
\end{eqnarray*}
and
\begin{equation*}
(-\nabla\Phi)^\rmr=\sum_{n=0}^{\infty}(n+1)\,r^{-(n+2)}\sum_{m=-n}^n a_{n,m}\,Y_n^m,
\end{equation*}
respectively.
The boundary condition \eref{eq:boundary1} then readily yields the coefficients as
\begin{eqnarray}
a_{1,0} &= b_z\,\sqrt{\pi/3}\,R^3\,\rmd z,\nonumber\\
a_{1,-1} &= -b_z\,\sqrt{\pi/24}\,R^3\,(\rmd x + \rmi\,\rmd y),\nonumber\\
a_{1,1} &= b_z\,\sqrt{\pi/24}\,R^3\,(\rmd x - \rmi\,\rmd y),\nonumber\\
a_{2,0} &= b_z\,\sqrt{4\pi/45}\,R^5,\label{eq:coeff1}
\end{eqnarray}
all other coefficients being zero.
We also introduce normalized quantities by measuring length in units of $R$ and the magnetic field in units of $b_z R$, i.e.
\begin{equation*}
\tix=\bi{x}/R,\,\,\tidx=\bi{\rmd x}/R,\,\,\tib(\tix)=\bi{B}(\tix R)/(b_z R).
\end{equation*}
Selected normalized field components are plotted in figure \ref{fig2}. As the field is zero inside the sphere and there is by definition no current outside the sphere, the supercurrent density $\bi{K}$ vanishes everywhere except on the surface of the sphere and is related to the transverse magnetic field by
\begin{equation*}
\bi{K}^{\theta}=-\bi{B}_\rmout^\mathrm{\Phi}|_{_{r=R}}/\mu_0,\;\;\bi{K}^{\phi}=\bi{B}_\rmout^\mathrm{\theta}|_{_{r=R}}/\mu_0,
\end{equation*}
where $\mu_0$ denotes the vacuum permeability. 
The normalized supercurrent distribution 
\begin{equation*}
\tik(\theta,\phi)=\mu_0\bi{K}(\theta,\phi)/(b_z R)
\end{equation*}
is thus obtained as
\begin{eqnarray}
\tik^{\theta}=&\case{3}{4}(\rmd\tilde{x}\,\sin\phi-\rmd\tilde{y}\,\cos\phi),\nonumber\\
\tik^{\phi}=&\case{3}{4}\cos\theta\,(\rmd\tilde{x}\,\cos\phi+\rmd\tilde{y}\,\sin\phi)\nonumber\\
&+\case{3}{2}\,\rmd\tilde{z}\,\sin\theta+\case{5}{4}\sin(2\theta).\label{eq:current1}
\end{eqnarray}

\section{Magnetic field for finite $\lambda$}
\begin{figure}[ht!]
\centering
\includegraphics[width=\linewidth]{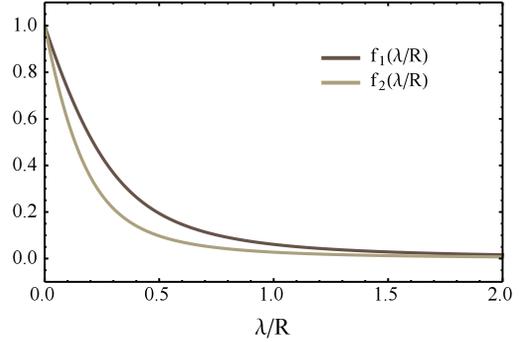}
\caption{Plot of the scaling functions $f_1$ and $f_2$. As shown in the next paragraph, the force on the sphere scales as $f_1$ with increasing $\lambda/R$.}
\label{fig3}
\end{figure}

For finite $\lambda$ the magnetic field inside the sphere is finite and determined by the London equation, i.e.
\begin{equation}
\label{eq:london1}
\Delta\bi{A}_\rmin = 1/\lambda^2\bi{A}_\rmin,\,\,\bi{B}_\rmin =\nabla\times\bi{A}_\rmin.
\end{equation}
The boundary condition that takes the place of \eref{eq:boundary1} is simply
\begin{equation}
\label{eq:boundary2}
\bi{B}_\rmin|_{_{r=R}}=\bi{B}_\rmout|_{_{r=R}}.
\end{equation}
To find the solution we first introduce the vector spherical harmonics\cite{0143-0807-6-4-014}
\begin{equation*}
\bi{Y}_{n}^{m} = Y_{n}^{m}\,\bi{e}_\rmr,\;\;
\bi{\Psi}_{n}^{m} = r\,\nabla Y_{n}^{m},\;\;
\bi{\Phi}_{n}^{m} = \bi{e}_\rmr\times\bi{\Psi}_{n}^{m}
\end{equation*}
and make the ansatz 
\begin{equation*}
\bi{A}_\rmin=\sum_{n=0}^{\infty}\sum_{m=-n}^n C_{n,m}(r)\,\bi{\Phi}_{n}^{m}(\theta,\phi).
\end{equation*}
Note that if a solution of this form with continuously differentiable $C_{n,m}(r)$ exists, then 
\begin{equation*}
\nabla\big(C_{n,m}(r)\,\bi{\Phi}_{n}^{m}(\theta,\phi)\big)=0
\end{equation*}
and the Coulomb gauge condition is fulfilled. The London equation \eref{eq:london1} leads to 
\begin{equation*}
r\,\partial_r^2\big(rC_{n,m}\big)-\big(r^2/\lambda^2+n(n+1)\big)C_{n,m}=0,
\end{equation*}
which is the modified spherical Bessel equation. The only solution (convergent at the origin) for the inside vector potential is therefore given by
\begin{equation*}
\bi{A}_\rmin=\sum_{n=0}^{\infty}i_n(\case{r}{\lambda})\sum_{m=-n}^n c_{n,m}\,\bi{\Phi}_{n}^{m}(\theta,\phi),
\end{equation*}
where the $i_n$ are first order modified spherical Bessel functions\cite{abramowitz+stegun} and the coefficients $c_{n,m}$ are yet to be determined. 
The fields can be expressed in vector spherical harmonics as
\begin{eqnarray}
\fl\bi{B}_\rmin=-\case{1}{r}\sum_{n=0}^{\infty}\sum_{m=-n}^n c_{n,m}\big[&n(n+1)\,i_n(\case{r}{\lambda})\,\bi{Y}_n^m\nonumber\\
&+\partial_r\big(r\,i_n(\case{r}{\lambda})\big)\,\bi{\Psi}_n^m\big],\label{eq:infield1}
\end{eqnarray}
\begin{equation*}
\bi{B}_\rmout=\bi{B}_0+\sum_{n=0}^{\infty}r^{-(n+2)}\sum_{m=-n}^n a'_{n,m}\big((n+1)\,\bi{Y}_n^m-\bi{\Psi}_n^m\big),
\end{equation*}
\begin{eqnarray*}
\bi{B}_0 = b_z\big[&-\sqrt{4\pi/3}\,\rmd z\,(\bi{Y}_1^0+\bi{\Psi}_1^0)\\
&+\sqrt{\pi/6}\,(\rmd x+\rmi\,\rmd y)\,(\bi{Y}_1^{-1}+\bi{\Psi}_1^{-1})\\
&-\sqrt{\pi/6}\,(\rmd x-\rmi\,\rmd y)\,(\bi{Y}_1^1+\bi{\Psi}_1^1)\\
&-\sqrt{\pi/5}\,r\,(2\,\bi{Y}_2^0+\bi{\Psi}_2^0)\big].
\end{eqnarray*}
The boundary condition \eref{eq:boundary2} then determines the coefficients as 
\begin{eqnarray*}
a'_{1,m} &= f_1(\case{\lambda}{R})\,a_{1,m},\\
a'_{2,0} &= f_2(\case{\lambda}{R})\,a_{2,0},\\
c_{1,m} &= (3\,\lambda)/\big(R^3\,i_2(\case{R}{\lambda})\big)\,a'_{1,m},\\
c_{2,0} &= \big(5\,\lambda\big)/\big(2\,R^4\,i_3(\case{R}{\lambda})\big)\,a'_{2,0},
\end{eqnarray*}
with 
\begin{eqnarray*}
f_1(\case{\lambda}{R})=1-3\,\case{\lambda}{R}(\coth\case{R}{\lambda}-\case{\lambda}{R}),\\
f_2(\case{\lambda}{R})=1-5\,\case{\lambda}{R}\big((\coth\case{R}{\lambda}-\case{\lambda}{R})^{-1}-3\,\case{\lambda}{R}\big).
\end{eqnarray*}
Here the $a'_{n,m}$ denote the coefficients of the scalar potential for finite $\lambda$, while the $a_{n,m}$ still refer to the coefficients for $\lambda=0$ as given in \eref{eq:coeff1}. Note that $a'_{n,m}$ is related to $a_{n,m}$ by a scaling function that depends only on the ratio $\lambda/R$. For $\lambda/R\rightarrow 0$ we get $a'_{n,m}\rightarrow a_{n,m}$ and $c_{n,m}\rightarrow 0$ and the solution for finite $\lambda$ thus converges to the solution for $\lambda=0$ determined in the last section. Normalized field components are plotted in figure \ref{fig2}, the scaling functions are plotted in figure \ref{fig3}. 
The supercurrent distribution $\bi{j}$ inside the sphere can now be simply obtained from the London equation \eref{eq:london1} and $\nabla\times\bi{B}_\rmin=\mu_0\bi{j}$ as $\bi{j}=-\case{1}{\mu_0\lambda^2}\bi{A}_\rmin$.

\section{Levitation force and maximum field}
The force on the sphere can be written in terms of the magnetic field on the surface of the sphere as
\begin{equation*}
\bi{F}=\case{R^2}{\mu_0}\int_0^{\pi}\rmd\theta\int_0^{2\pi}\rmd\phi\,\sin\theta\,\big[(\bi{e}_\rmr\bi{B})\bi{B}-\case{1}{2}\bi{e}_\rmr\bi{B}^2\big]_{_{r=R}}.
\end{equation*}
Here $\bi{B}$ denotes either $\bi{B}_\rmin$ or $\bi{B}_\rmout$, as they coincide on the surface of the sphere. Carrying out the integration we find 
\begin{equation*}
F_z = -\case{3V}{2\mu_0}\,b_z^2\,f_1(\case{\lambda}{R})\,\rmd z,\;\;\case{1}{\rmd x}F_x = \case{1}{\rmd y}F_y = \case{1}{4\,\rmd z}F_z,
\end{equation*}
where $V=\case{4\pi}{3}R^3$ is the volume of the sphere. For $\lambda \rightarrow 0$ we have $f_1(\case{\lambda}{R})\rightarrow 1$ and we recover the expressions given in \cite{PhysRevLett.109.147205}.

\begin{figure}[ht!]
\centering
\includegraphics[width=\linewidth]{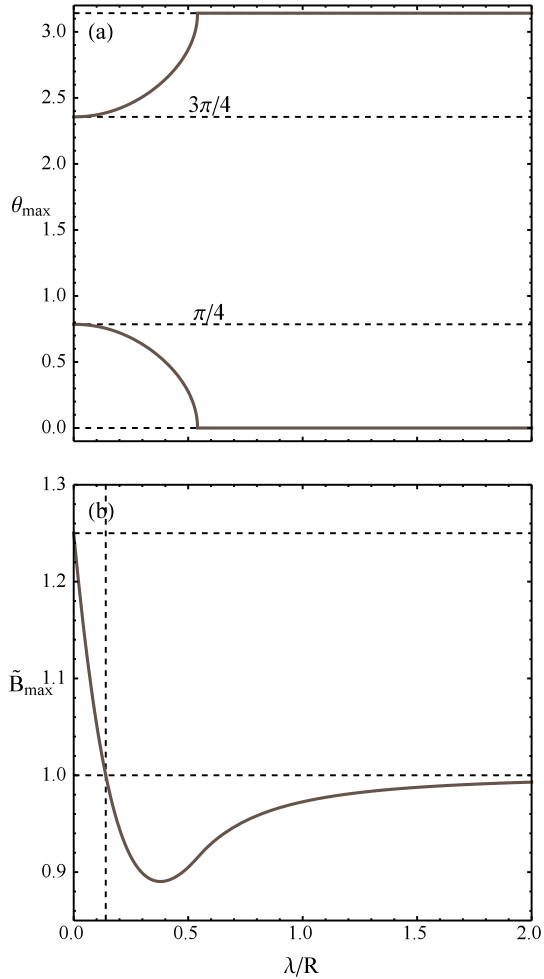}
\caption{(a)Plot of the double-valued angle for which the field strength is at a maximum. (b)Plot of the normalized maximum field strength $\tilde{B}_\mathrm{max}$. The dashed lines are guides to the eye.}
\label{fig4}
\end{figure}

We now determine the maximum field strength $B_\mathrm{max}=\mathrm{max}(\sqrt{\bi{B}_\rmin^2})$, where the maximum is evaluated over the volume of the sphere, for $\bi{\rmd x}=0$. In this case the squared magnetic field inside the sphere reduces to 
\begin{eqnarray*}
\fl\bi{B}_\rmin^2=&(45\,c_{2,0}^2)/(4\pi r^2)\big[(3\,\cos^2\theta-1)^2\,i_2^2(\case{r}{\lambda})\\
&+\cos^2\theta\,\sin^2\theta\,\big(\case{r}{\lambda}i_3(\case{r}{\lambda})+3\,i_2(\case{r}{\lambda})\big)^2\big].
\end{eqnarray*}
Evaluating the partial derivatives with respect to $r$ and $\theta$ it follows that $\partial_r\vec{B}_\rmin^2\geq 0\;\forall (r,\theta)$, i.e. the maximum lies on the surface of the sphere, and that the maximum occurs for $\theta=\theta_\mathrm{max}$ with $\theta_\mathrm{max}$ determined by 
\begin{equation}
\label{eq:maxangle}
\cos^2\theta_\mathrm{max}=\frac{\big(\case{R}{\lambda}i_3(\case{R}{\lambda})/i_2(\case{R}{\lambda})-3\big)^2-6}{2\Big(\big(\case{R}{\lambda}i_3(\case{R}{\lambda})/i_2(\case{R}{\lambda})-3\big)^2-9\Big)}
\end{equation}
for $\lambda/R\;\raisebox{1pt}{\scaleobj{0.8}{\lesssim}}\;0.54$ and $\theta_\mathrm{max}\in\{0,\pi\}$ otherwise. We then get $B_\mathrm{max}=|\bi{B}_\rmin(R,\theta_\mathrm{max})|$.
Note that the expression on the right hand side of \eref{eq:maxangle} converges to $\case{1}{2}$ for $\lambda/R\rightarrow 0$, which corresponds to $\theta_\mathrm{max} \rightarrow \case{\pi}{2} \pm \case{\pi}{4}$ and $B_\mathrm{max}\rightarrow\case{5}{4}b_zR$ (this result can of course also directly be read off equation \eref{eq:current1} derived in section 2 for $\lambda=0$). On the other hand, for $\lambda/R\rightarrow \infty$, i.e. a non-magnetic sphere, we simply have $\theta_\mathrm{max}\in\{0,\pi\}$ and $B_\mathrm{max}\rightarrow b_zR$. In figure \ref{fig4} we plot $\theta_\mathrm{max}$ as well as $\tilde{B}_\mathrm{max}$ against the ratio $\lambda/R$. It is interesting to note that for $\lambda/R\;\raisebox{1pt}{\scaleobj{0.8}{\gtrsim}}\;0.14$ the maximum field strength is smaller than $b_zR$, i.e. in that range $B_\mathrm{max}$ is actually smaller for a superconducting sphere than it would be for a non-magnetic sphere. This result is counter-intuitive at first glance and in stark contrast to the case of a superconducting sphere in a homogeneous field\cite{91920b12e27e4ccf8b445bddeb5b53c6} $\bi{B}_\mathrm{hom}$, where the maximum field strength for any value of $\lambda/R$ will always be higher than $|\bi{B}_\mathrm{hom}|$. One can understand this behavior qualitatively by looking at the absolute fields at $\theta=0$ and $\theta=\case{\pi}{4}$ for increasing values of $\lambda/R$. In the former case we have $|\bi{B}_\rmin(R,0,\phi)|_{_{\lambda=0}}=0$ monotonically increasing to $|\bi{B}_\rmin(R,0,\phi)|_{_{\lambda\rightarrow\infty}}\rightarrow b_zR$, while in the latter case we have $|\bi{B}_\rmin(R,\pi/4,\phi)|_{_{\lambda=0}}=\case{5}{4}b_zR$ monotonically decreasing to $|\bi{B}_\rmin(R,\pi/4,\phi)|_{_{\lambda\rightarrow\infty}}\rightarrow \sqrt{\case{5}{8}}b_zR$. Thus, when $\theta_\mathrm{max}$ shifts towards $\theta=0$ with increasing $\lambda/R$ we will, at some point, have $|\bi{B}_\rmin(R,\theta_\mathrm{max},\phi)|<b_zR$. 
Analytic solutions for the maximum field strength for $|\bi{\rmd x}|>0$ can be found as well, but the resulting expressions are bulky and do not serve to further enhance understanding of the physics. For anyone interested in these results we recommend starting with \eref{eq:infield1} and using a computer algebra system to derive the expressions for the maximum.

\section{Summary and discussion}
In the previous sections we have derived analytical solutions for the magnetic field distribution for a superconducting sphere in the Meissner state placed in an applied quadrupole field by analytically solving the Maxwell-London equations with appropriate boundary conditions. The solutions are obtained by expanding the fields in terms of vector spherical harmonics. We then derived the force acting on the sphere. The results are valid as long as the maximum field strength on the surface of the sphere is below a critical field strength $B_\mathrm{c1}$, the exact value of which depends on the superconducting material. Above $B_\mathrm{c1}$ the superconductor will enter an intermediate state (type-I) or mixed state (type-II)\cite{tinkham1996introduction}, respectively, and the analysis provided here can no longer be applied.
We also calculated the maximum field strength seen by the superconducting sphere when it is located at the center of the quadrupole field, and demonstrated that for a certain range of $\lambda/R$ the maximum field strength is lower than it is for a non-magnetic sphere. 

We expect these results to be applied in the context of quasi-static magnetic traps for superconductors and to greatly enhance understanding of these traps. From the analytic solutions for the force and the field distribution one can directly obtain analytic results for trapping frequencies and coupling strengths. Previous analysis was limited to spheres that are large compared to their penetration depth ($\lambda/R\rightarrow 0$), while our results are valid for arbitrarily sized spheres. 
Our results furthermore show a way to connect dynamical parameters of the magnetic trap (e.g. frequency) to material constants (e.g penetration depth), opening up new ways to measure these quantities. 

The solution can easily be extended to magnetic fields of various forms, as long as they possess an expansion in vector spherical harmonics.

\ack
The authors are grateful for discussions with A. Sanchez and his group. This work was supported by the European Union‘s Horizon 2020 research and innovation programme under grant agreement No 736943 (MaQSens), the European Research Council (ERC CoG QLev4G), and the Austrian Science Fund (FWF) under project F40 (SFB FOQUS).

\section*{References}
\bibliography{iopart-num}

\end{document}